\documentclass[aps,prl,floatfix]{revtex4-1}
\usepackage{graphicx}
\usepackage{bm}
\usepackage{dsfont}
\usepackage{color}

\newcommand{\ignore}[1]{}

\newcommand{\be}{\begin{equation}}
\newcommand{\ee}{\end{equation}}
\newcommand{\bea}{\begin{eqnarray}}
\newcommand{\eea}{\end{eqnarray}}

\newcommand{\ket}[1]{\left\vert #1    \right\rangle }
\newcommand{\bra}[1]{\left\langle   #1  \right\vert}

\begin{document}

\title{Quantum Gates and Memory using Microwave Dressed States}

\author{N. Timoney}
\author{I. Baumgart}
\author{M. Johanning}
\author{A. F. Var\'{o}n}
\author{Ch. Wunderlich}
\affiliation{Department Physik, Naturwissenschaftlich-Technische Fakult\"{a}t,  Universit\"{a}t Siegen, 57068 Siegen, Germany}

\author{M. B. Plenio}
\author{A. Retzker}
\affiliation{Institut f{\"u}r Theoretische Physik, Universit{\"a}t Ulm, 89069 Ulm, Germany}

\date{\today}
\begin{abstract}
Trapped atomic ions have been successfully used for demonstrating
basic elements of universal quantum information processing (QIP)
\cite{Blatt2008}. Nevertheless, scaling up of these methods and
techniques  to achieve large scale universal QIP, or more
specialized quantum simulations
\cite{Friedenauer2008,Kim2010,Gerritsma2010,Johanning2009} remains
challenging. The use of easily controllable and stable microwave
sources instead of complex laser systems
\cite{Mintert2001,Ospelkaus2008} on the other hand promises to
remove obstacles to scalability. Important remaining drawbacks in
this approach are the use of magnetic field sensitive states, which
shorten coherence times considerably, and the requirement to create
large stable magnetic field gradients. Here, we present
theoretically a novel approach based on dressing magnetic field
sensitive states with microwave fields which addresses both issues
and permits fast quantum logic. We experimentally demonstrate basic
building blocks of this scheme to show that these dressed states are
long-lived and coherence times are increased by more than two orders
of magnitude compared to bare magnetic field sensitive states. This
changes decisively the prospect of microwave-driven ion trap QIP and
offers a new route to extend coherence times for all systems that
suffer from magnetic noise such as neutral atoms, NV-centres,
quantum dots, or circuit-QED systems.
\end{abstract}
\maketitle

{\em Introduction --- } Using laser light for coherent manipulation
of qubits gives rise to fundamental issues, notably, unavoidable
spontaneous emission which destroys quantum coherence
\cite{Ozeri2007,Plenio1997}. The difficulty in cooling a collection
of ions to their motional ground state and the time needed for such
a process in the presence of spurious heating of Coulomb crystals
limits the fidelity of quantum logic operations in laser-based
quantum gates, and thus hampers scalability. This limitation is only partially removed by the use of 'hot' gates
\cite{Sorensen2000,Milburn2000}.
Technical challenges in accurately controlling the frequency and intensity of laser light as well as delivering 
a large number of laser beams of high intensity to trapped ions are further obstacles for scalability.

These issues
associated with the use of laser light for scalable QIP
have lead to the development of novel concepts for performing
conditional quantum dynamics with trapped ions that rely on radio frequency (rf) or
microwave (mw) radiation instead of laser light
\cite{Mintert2001,Wunderlich2002,Wunderlich2003,McHugh2005,Ospelkaus2008,Wang2009}.
Rf or mw radiation can be employed for quantum gates through the
use of magnetic gradient induced coupling (MAGIC) between spin
states of ions
\cite{Johanning2009a},
thus averting  technical and fundamental issues of scalability that
were described above. Furthermore, the sensitivity to motional
excitation of ions is reduced in such schemes. A drawback
of MAGIC is the necessity to use magnetic field sensitive states for
conditional quantum dynamics, thus making qubits susceptible to
ambient field noise and shortening their coherence time. This issue
is shared with some optical ion trap schemes for QIP that usually rely on
magnetic field sensitive states for conditional quantum dynamics, like geometric gates
\cite{Blatt2008}, limiting the coherence time of qubit states
typically to a few ms. In an effort to extend the coherence time of
atomic states, two-qubit entangled states forming a decoherence-free
subspace have been created \cite{Haeffner2005a,Kielpinski2001}.
Recently, transfer between field sensitive states, that are used for
conditional quantum dynamics and field insensitive states used for
storage of quantum information has been employed \cite{Home2009a}.

The relevant noise source in this case, namely magnetic field fluctuations,
is not featureless white noise but tends to have a limited bandwidth.
In this context techniques were proposed for prolonging coherence times
by subjecting the system to a rapid succession of pulses leading to a
decoupling from the environment. This technique, termed Bang Bang control
\cite{Viola1998}, and its continuous version\cite{Rabl2009} can be applied to advantage
in a variety of systems including hybrid atomic and nano-physics technologies. Recent work includes the experimental demonstration
of optimized pulse sequences made for suppression of qubit decoherence
(\cite{Biercuk2009,Bluhm2011} and references therein).

Here, we encode
qubits in microwave dressed states requiring only continuous,
constant intensity microwave fields. This scheme protects qubits
from magnetic field fluctuations and, importantly, allows at the
same time to perform fast quantum gates even for small Lamb-Dicke
parameters, and therefore moderate magnetic field gradients.
Microwave generating elements for coherent manipulation of qubits
can be integrated in micro-structured ion traps
\cite{Ospelkaus2008} such that quantum information
processing can be realized using scalable ion chips.
Thus, this novel scheme is a significant step
on the route towards integrating elements required for quantum
information prcoessing on a scalable ion chip. Moreover, the ideas
presented here are generic and can be applied to all laser- or
microwave-based QIP such as neutral atoms, NV-centres and quantum
dots.

We describe theoretically the scheme for storage, single and
multi-qubit quantum gate operation and then present experimental
demonstrations of storage and information processing of quantum
information that demonstrate gains of two orders of magnitude in
coherence times.

\begin{widetext}
\begin{figure}
\vspace*{-0.8cm}
\includegraphics[width=0.95\textwidth]{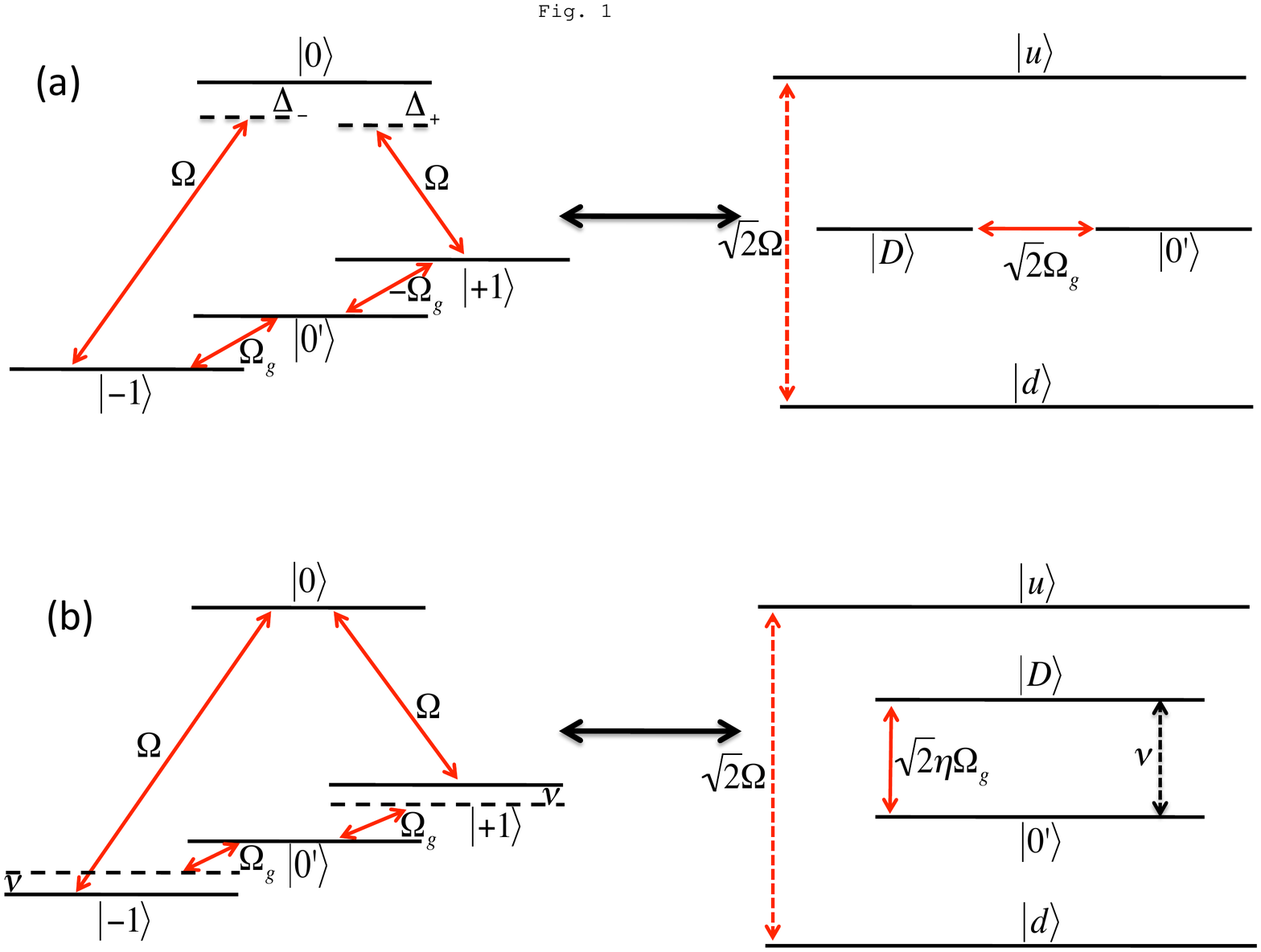}
\caption{
Microwave-dressed qubit states.
$\ket{-1}$ and $\ket{+1}$ represent internal magnetic sensitive
states, e.g. $m_F=\pm 1$, and $\ket 0$ and $\ket{0'}$ represent two additional
magnetically insensitive levels (e.g. $m_F=0$ states of two different hyperfine
manifolds). The application of microwave radiation, with Rabi frequency $\Omega$,
induces a dressed-state basis
where the states $\ket D = (\ket{-1}-\ket{+1})/\sqrt{2}$ and $|0'\rangle$ are
separated by an energy gap from the other two states, $\ket u = (\ket B +
\ket 0 )/\sqrt{2}$ and $\ket d = (\ket B - \ket 0 )/\sqrt{2}$, where $\ket B
= (\ket {-1}+\ket{+1})/\sqrt{2}$. This arrangement creates a qubit in
$\ket D$ and $|0'\rangle$ that is resilient to magnetic field fluctuations --
since phase changes are now suppressed by an energy gap -- and to microwave power
fluctuations, since the energy gap between $\ket D$ and $\ket {0'}$ does not
depend on microwave power.
As shown in part (a), an rf field with Rabi frequency $\Omega_g$ and detuning
$\Delta_g$, which should be equal to $\nu$ for a multi-qubit gate and should be zero for a single-qubit gate, can be used to implement general single qubit quantum gates. The right part is the the frame which rotates with the driving frequencies (dressed states).
Conditional quantum dynamics coupling electronic degrees of freedom to the
motion can be achieved (see (b)) in the presence of a magnetic field gradient
when an rf field that is detuned from the carrier transition of the qubit
by the vibrational frequency $\nu$. In this case the $\ket D$ and $\ket {0'}$
are not degenerate but have an energy gap which is the rf detuning.
The atomic hyperfine states of the electronic ground state of $^{171}$Yb$^+$ used in this work can be mapped to the
general scheme shown here as follows: $\ket{F=0} \leftrightarrow \ket{0}$,
$\ket{F=1, m_F=-1} \leftrightarrow \ket{-1}$, $\ket{F=1, m_F=+1}
\leftrightarrow \ket{+1}$, $\ket{F=1, m_F=0} \leftrightarrow \ket{0'}$
}
\label{SingleQubit}
\end{figure}
\end{widetext}

{\em Theory of dressed state memory and single-qubit ---} We
consider a typical energy level configuration depicted in Fig. 1a.
Encoding quantum information in the subspace
spanned by two magnetically sensitive $m_F=\pm 1$ states 
($\ket{-1}, \ket{+1}$) will lead to a rapid loss of coherence due to
fluctuating magnetic fields.
We first note that microwave-dressed states (with Rabi frequency
$\Omega$ and detunings $\Delta_-$ and
$\Delta_+$ which are optimally equal) create a subspace spanned by the two states $|D\rangle =
(\ket{-1}-\ket{+1})/\sqrt{2}$ and  $|0'\rangle$  ($m_F=0$) of
the same manifold that is separated from the other eigenstates
$|u\rangle$ and $|d\rangle$ of the system by a finite energy gap.
The energetically degenerate states $|D\rangle$ and $|0'\rangle$ are
not coupled by magnetic field fluctuations. As a consequence of the
energy gap to the states $|u\rangle$ and $|d\rangle$, a relative
phase change between states $\ket{-1}$ and $\ket{+1}$ now acquires
an energy penalty and dephasing is strongly suppressed as long as
the spectral power density of the magnetic field fluctuations at the
frequency corresponding to the energy gap is negligible.
Note however that the levels $|D\rangle$ and $|0'\rangle$ are
energetically degenerate independent of the applied microwave
Rabi-frequency $\Omega$ and hence stable against its fluctuations.
Hence we will consider the qubit encoded in the subspace spanned by
$|D\rangle$ and $|0'\rangle$. In this case, the dephasing time would
be limited by second order effects in the magnetic field
fluctuations which in our case leads to a lifetime of the order 1 s.
The nature of the protected subspace depends on the relative phase
between the microwave fields on the transitions $|0\rangle
\leftrightarrow |\pm 1\rangle$. The above description applies for a
relative phase $0$ while for a relative phase of $\pi$, the protected
subspace is spanned by $|B\rangle = (\ket{-1}+\ket{+1})/\sqrt{2}$
and $|0'\rangle$. We have implemented both settings in the
experiments reported below.

The application of additional rf fields (with Rabi frequency
$\Omega_g$ and possible detuning $\Delta_g$) and relative phase
$\pi$ then permits the implementation of general single qubit
rotations in the protected subspace spanned by $|D\rangle$ and
$|0'\rangle$. Choosing $\Delta_g=0$ yields the Hamiltonian $H =
\Omega_g \left(\ket D \bra{0'} +\ket{0'}\bra D \right)$ which
describes arbitrary rotations about the $x-axis$ in the protected space.
A rotation about the z-axis is
obtained for $\Omega_g, \nu\ll \Delta_g\neq 0$. The specific case $\Delta_g=\nu$, i.e.
tuning to the motional sideband will be discussed later as it couples the electronic
to the motional degree of freedom.

{\em Experimental realization --}
Hyperfine states of electrodynamically trapped $^{171}$Yb$^+$ ions characterized
by quantum numbers $F=0$ and $F=1$ in the electronic ground state S$_{1/2}$ are
used in the experiments reported here (see the caption of Fig. 1).
State-selective detection is achieved by collecting scattered fluorescence on
the ($S_{1/2}, F=1) - (P_{1/2}, F=0$)  resonance which allows for discriminating
population in states $S_{1/2}, F=0$ (no fluorescence) and $S_{1/2}, F=1$  (resonance fluorescence is detected).

The procedure to generate and detect dressed states can be thought
of as being divided into three segments (compare inset of Fig. 2):
First (up to time $t=T_1$), an incomplete stimulated rapid adiabatic
passage (STIRAP) \cite{Vitanov2001} is used to adiabatically
transfer the system from the atomic to the dressed state basis.
Second (up to $t=T_2$), the amplitudes of the dressing fields are kept
constant for a holding time $T=T_2 - T_1$, since we are interested in creating
and  utilising dressed states (as opposed to population transfer).
Between $t=T_1$ and $t=T_2$ dressed states are present (right-hand-side of Figs.1a and 1b) and quantum operations with them 
are implemented, for example,
through application of additional rf fields. For a single-qubit gate between dressed states
(Fig. 1a rhs), the rf field is on resonance with the
$\ket{0'}\leftrightarrow \ket{+1} ,\ket{-1}$ transition,
and for a multi-qubit gate (Fig. 1b) the rf is
detuned by the vibrational mode frequency $\nu$ from this transition. 
Third ($t>T_2$), the STIRAP sequence is
completed when the ion is transferred from the dressed state basis
back to the atomic states. Any dephasing of a dressed state or
transitions to other states during the holding time $T$ gives rise
to imperfect population transfer during step three of this sequence
that returns the system to the atomic state basis. This is described
in more detail in the Methods section for the creation and detection
of the $\ket D$ state.

\begin{figure}
\begin{center}
\includegraphics[height=0.3\textheight]{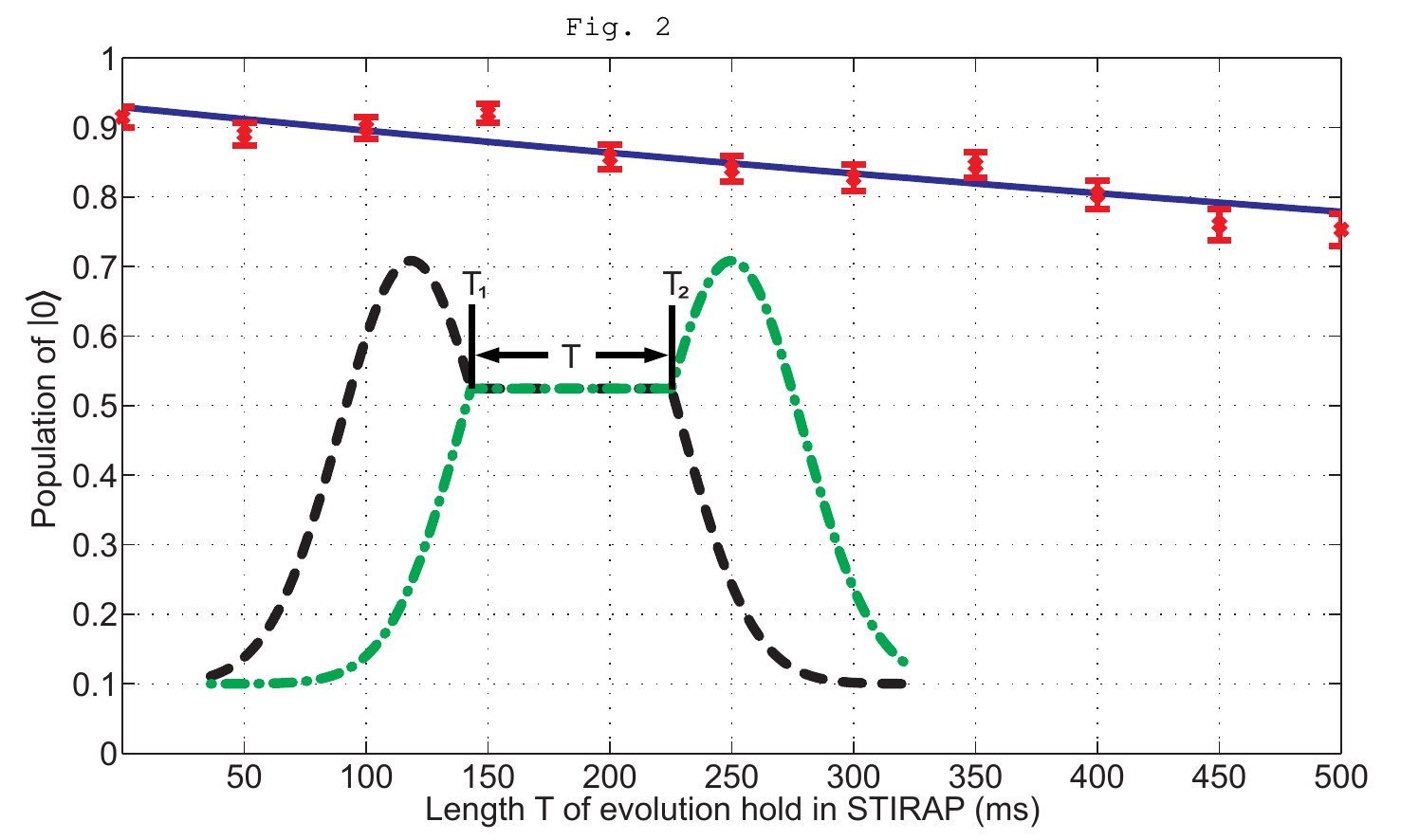}
\end{center}
\caption{
Lifetime of the dressed state $\ket{D}$.
The inset shows the qualitative temporal behaviour of the two dressing fields' amplitudes. 
In order to prepare the dressed state $\ket{D}$, the dressing field's amplitudes are ramped up adiabatically until time $T_1$. Then, the amplitudes of these dressing fields are kept constant until time $T_2$.  Between $T_1$ and $T_2$ the dressed states shown in Fig. 1 (right-hand-side) are present. Finally, after time $T_2$, the STIRAP sequence is completed adiabatically transferring the dressed state $\ket{D}$ back to a bare ionic state. The hold time $T = T_2 - T_1$. 
The solid line represents a fitted exponential giving a lifetime of $1700 \pm 300$ ms. 
Both microwave dressing fields are
set on resonance with the atomic transitions (in the bare state
picture they drive the $\ket{-1} - \ket{0}$ and the $\ket{+1} -
\ket{0}$ transitions). The adiabatic (STIRAP) preparation is
characterized by these parameters: pulse separation of
$6/f_{\Omega}$, pulse width of $5/f_{\Omega}$
and $\Delta t = \frac{1}{10f_{\Omega}}$, where
$f_\Omega=\Omega/(2\pi) = 36.5$ kHz. The microwave frequency on the $\ket{+1} - \ket{0}$ resonance was 12.6528121 GHz, and on the on the $\ket{-1} - \ket{0}$ resonance it was 12.6328272 GHz; a static
magnetic field $B=0.714 $ mT defines a quantization axis. Each measurement point
consists of 300 repetitions.  
}
\label{hold}
\end{figure}

{\em Preparation and lifetime of state $|D\rangle$ --} Fig. 2
presents the effectiveness of preparation and detection
of state $\ket D$ as a function of the holding time $T$. The
measurement shown there extends over 500 ms and an exponential fit of the data yields a
lifetime of
this state of 1700 $\pm 300$ ms.
This lifetime is limited by magnetic field fluctuations
that couple the $\ket D$ state to other states and the creation of a
different dark state through phase fluctuations between the two
microwave fields.
This represents a remarkable improvement by more than
two orders of magnitude in the dependance on the magnetic
fluctuations, as the dephasing time of magnetically sensitive bare
atomic states $|\pm1\rangle$ in this apparatus has a measured
coherence time not exceeding 5.3 ms.

We have investigated the effectiveness of the STIRAP process
for creating dressed states and transferring them into the final
state as a function of the parameters characterizing the pulse
sequence and found this technique to be
robust over a wide range of experimental parameters (see Methods).

{\em Single qubit gates and coherence times of dressed state qubit
---} As explained, our scheme permits the implementation of general single-qubit rotations in the subspace spanned by $\{ \ket{0'}, \ket D \}$
whose states are resilient not only to magnetic field fluctuations
but also to variations in the amplitude of the dressing fields.
The reason for this is that magnetic fluctuations couple the $\ket{B}$ to
the $\ket{D}$ state and since the magnetic field noise power spectrum is very
small at the microwave driving frequency their effect is substantially reduced. 
The new dephasing time will be governed by second order corrections which 
will scale with $\left(\frac {\Delta B}{\Omega}\right)^2$ with respect to the original dephasing rate, where $\Delta B$ is the amplitude of the fluctuations.
Moreover, the microwave does not couple any states in the qubit subspace and thus does not limit the phase coherence.

\begin{figure}
\vspace*{-2.5cm}
\includegraphics[height=0.5\textheight]{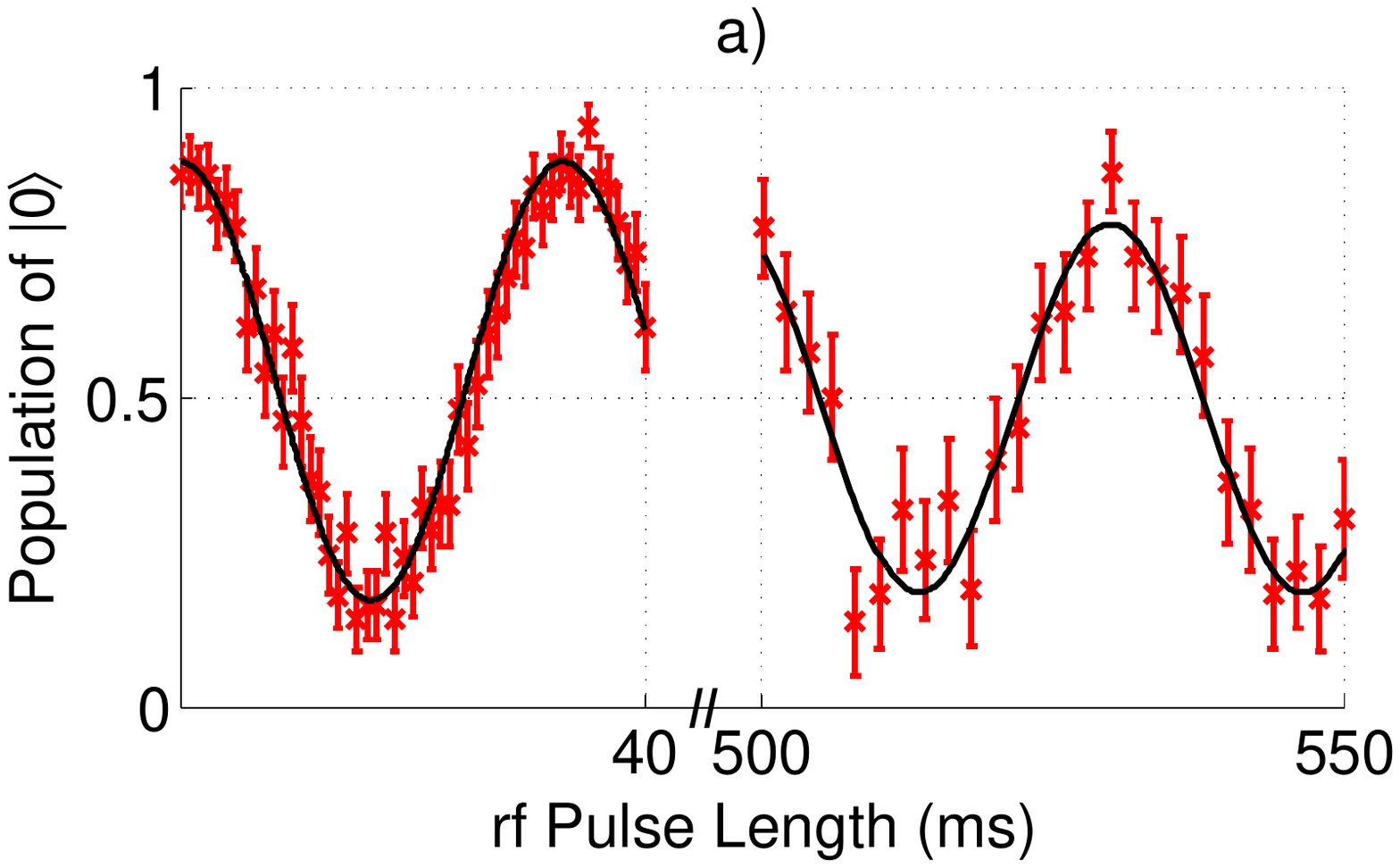}
\includegraphics[height=0.5\textheight]{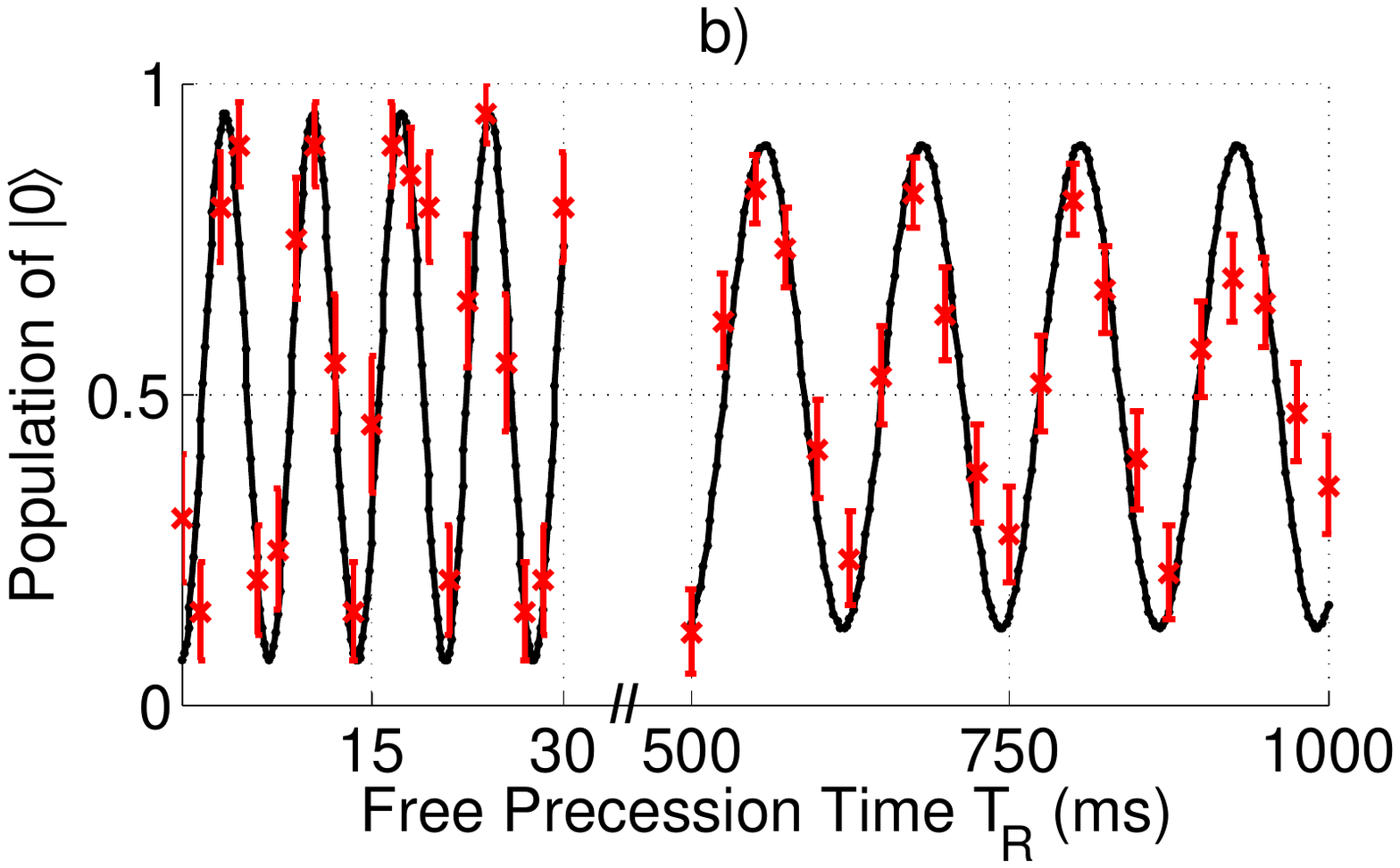}
\vspace*{-2.5cm}
\caption{
Single-qubit gates with dressed states.
a) Rabi oscillations between dressed
state $\ket{B}$ and $\ket{0'}$ induced by an rf field for times up
to more than 500 ms. Population in $\ket{B}$ is mapped onto state
$\ket{0}$ at the end of the STIRAP and detection sequence (each
datapoint is the average of 50 [up to 40 ms] or 25 repetitions [over 500 ms]).
The microwave frequency on the $\ket{+1} - \ket{0}$ resonance was 12.6533088 GHz, and on the $\ket{-1} - \ket{0}$ resonance it was 12.6323327 GHz; the microwave Rabifrequency $\Omega = 31.8 \times 2 \pi$ kHz; the rf frequency driving transitions between dressed states was set to 10.49676 MHz; a static
magnetic field $B=0.749 $ mT defines a quantization axis.
b) Ramsey-type measurement  preparing a
coherent superposition of $\ket{B}$ and $\ket{0'}$ and probing it
after time $T_R$. Two rf $\pi/2$-pulses separated by time $T_R$ of
free evolution are applied to the qubit transition. The rf is
slightly detuned from resonance (near 10.265 MHz corresponding to the $\ket{-1} - \ket{0}$ and $\ket{+1} - \ket{0}$ transitions) yielding Ramsey fringes with period $1/(144.4 Hz)$ between 0.1 ms and 30 ms. For the measurement between 500 ms  and 1000 ms the period is $1/ (8.069 Hz)$
(0 ms -30 ms: 20 repetitions per datapoint; 500 ms - 1000 ms: 40 repetitions).
Here, $f_\Omega=\Omega/(2\pi) = 37.3$ kHz. The microwave frequency on the $\ket{+1} - \ket{0}$ resonance was 12.6530938 GHz, and on the on the $\ket{-1} - \ket{0} $resonance it was 12.6325472 GHz; a static magnetic field $B=0.730 $ mT defines a quantization axis.
}
\label{fig:gate}
\end{figure}

In order to demonstrate the enhanced coherence time of this qubit,
we have conducted Rabi- and
Ramsey-type measurements. First we move the system to the dressed
state basis and then apply additional rf-fields with Rabi-frequency
$\Omega_g$ that induce Rabi oscillations between the dressed states
$|B\rangle$ and $|0'\rangle$. Using the $|B\rangle$ state (not
subject to spontaneous decay) has the advantage that  no phase of
$\pi$ between the rf-fields is needed. Completing the STIRAP cycle
then maps the system to an atomic state that depends on the position
of the atom in the dressed state Rabi-cycle. Results that we have
obtained for this procedure are shown in Fig. 3a.
The Rabi oscillations are sustained over 550 ms demonstrating the
long-lived coherence of the dressed states when driven by rf
radiation.

These experiments  also demonstrate that coherent transfer to
the dressed state basis and subsequent application of a Rabi pulse prepares
a coherent superposition in the qubit space that can then be read out efficiently
after completion of the STIRAP cycle. This in turn forms the basis for the
measurements of the lifetime of coherent superpositions in the protected
subspace that we are now turning to.

These Ramsey-type experiments
test the dephasing time of the dressed state qubit. Fig. 3b
shows that coherence is preserved for more
than 1000 ms,
close to the ultimate limit of about 1700 ms set by the
lifetime of the state $|D\rangle$ and more than two orders of
magnitude longer than the dephasing time of the atomic states
$\ket{-1}$ and $\ket{+1}$ making this scheme ideal for realizing a
quantum memory.

{\em Coupling qubit to motion --- } The realization of multi-qubit gates can be
achieved by coupling the electronic qubit to the motion of the ions \cite{Cirac1995,Sorensen2000,Milburn2000}. Several schemes
for realizing such conditional quantum dynamics are possible using dressed states.
Here we outline the scheme illustrated in Fig. 1b.

This can be achieved by using a pair of rf-fields on the $|0'\rangle
\leftrightarrow |\pm1\rangle$ transitions. These rf fields are detuned by the vibrational mode frequency $\nu$
from this carrier and are thus in resonance with the first motional
side band (compare Fig. 1b).
In the dressed state basis, this couples the $\ket D \leftrightarrow \ket{0'}$ 
qubit resonance of the protected subspace to the vibrational 
mode of the ion string. The coupling $\ket B \leftrightarrow \ket{0'}$ in zeroth
order in the Lamb-Dicke parameter (carrier transition) will be
canceled in the rotating wave approximation (RWA) due to the higher 
energy (by $\Omega$) of the $\ket B$ state, and state $\ket D$ will be coupled
only in first order in the Lamb Dicke parameter such that we obtain
a Hamiltonian of the form (see Methods for details) \be
    H = \sqrt{2}\eta \Omega _g\left( \left\vert D\right\rangle \left\langle
    0'\right\vert e^{i\delta t}+h.c\right) (b^{+}-b).
\ee 
Importantly, since this gate has no carrier part, a small
Lamb-Dicke parameter $\eta$ can be compensated for by increasing the rf
power while obeying $\eta \Omega_g \ll \nu$. This will allow for working with moderate static
\cite{Mintert2001,Wunderlich2002,Johanning2009} or oscillating
\cite{Ospelkaus2008} magnetic field gradients when utilizing rf or
microwave radiation for multi-qubit quantum gates. Also, this scheme
permits the realization of other types of gates including gates with
ions in thermal motion
\cite{Sorensen2000,Milburn2000}.

{\em Multi-qubit quantum processor --- } To complete the elements of
a universal QIP device based on microwave dressed states we will now
outline how the dressed state scheme for QIP can be applied to a collection
of trapped ions. The long wavelength of the microwave radiation requires
the use of a static magnetic field gradient, if individual addressing in frequency space is desired.
A gradient of a static or oscillating magnetic field is required for the generation of the coupling of
the electronic degrees of freedom to the motional degrees of freedom when microwave
or rf radiation is used for quantum gates. Therefore, this does not represent an additional experimental requirement.

\begin{widetext}
\begin{figure}
\includegraphics[width=0.5\textwidth]{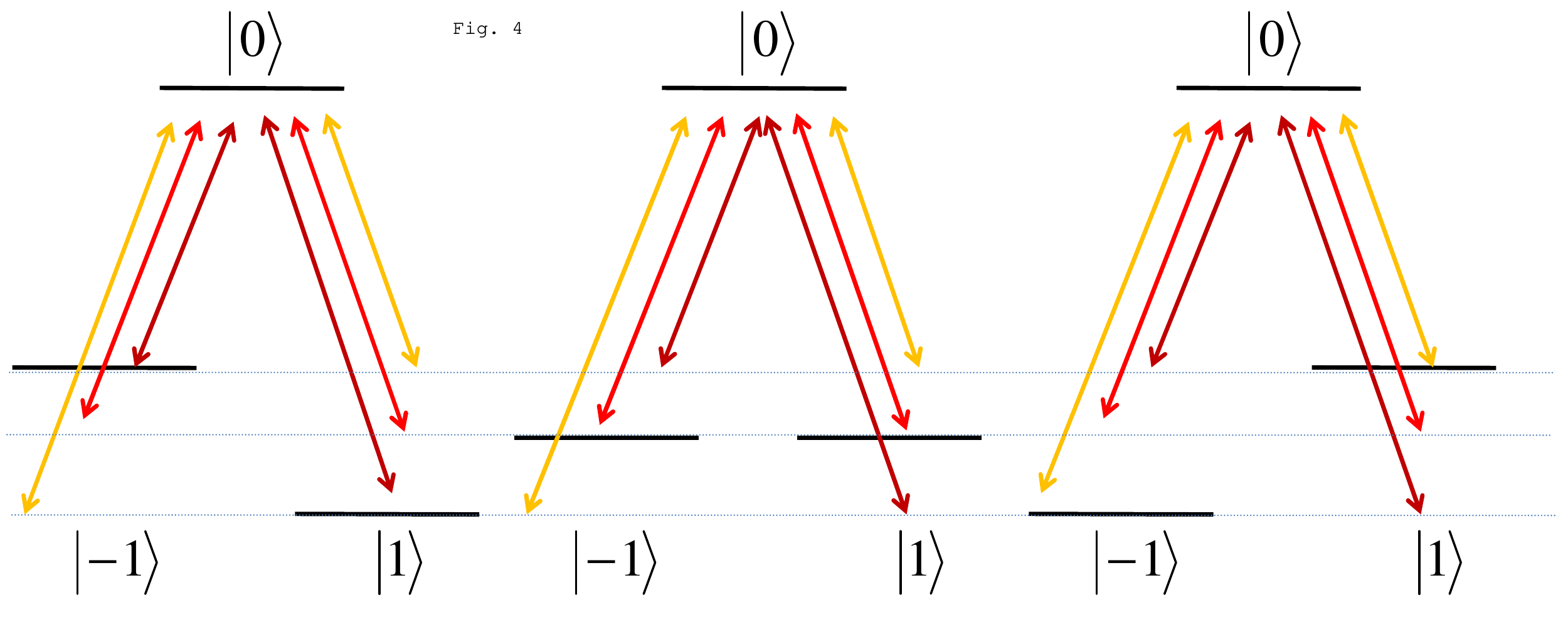}
\caption{
Multi-qubit scheme.
Energy level representation of three ions along a gradient
of magnetic field. The magnetic field on the two external ions has
the same magnitude but opposite direction producing opposite
symmetrical Zeeman energy shifts on the levels $\ket{-1}$ and
$\ket{1}$. The arrows represent three pairs of microwave frequencies
from a comb with frequency interval corresponding to the Zeeman
shift between two ions. Each frequency pair (identified by its own
color) is only resonant with one ion.
}
 \label{system}
\end{figure}
\end{widetext}

In the absence of a static magnetic field gradient, a single pair of
microwave dressing fields is sufficient to dress all ions. In the
presence of static gradient each ion would require its own pair
of driving fields. This can be accomplished efficiently by employing
a microwave frequency comb where the frequency spacing coincides
with the change in Zeemann shift between neighbouring ions as shown
in Fig. 4.
Then, for each ion two microwave fields in the
comb are resonant with the relevant atomic transition, dressing and
shifting the states $\ket u $ and $\ket d $. For each off-resonant
microwave field with detuning $\Delta$ there is a second field with
equal and opposite detuning $-\Delta$. Hence each such off-resonant
pair will only couple to $\ket B$ leading to equal and opposite
Stark-shifts which then cancel. Thus the dressed state structure for
each ion remains essentially the same as in the single ion case
achieving a robust memory. Small errors due to higher order effects
may be introduced and the execution of gates on ions may lead to
off-resonant transitions and hence phase shifts in other ions.
However, it is important to note that the amount of this error is
exactly as for MAGIC
\cite{Mintert2001}. This is due to the fact that the microwave
fields almost completely decouple the $\{\ket u, \ket d\}$ subspace
from the qubit subspace($\{\ket D, \ket {0'}\}$) as long as $\Omega$ (the dressing field Rabi
frequency) is much larger (typically, an order of magnitude) than
the rf coupling $\Omega_g$. The reason for this decoupling is the large energy gap between the qubit subspace and the $\{\ket u, \ket d\}$ subspace.
 For both single qubit gates and
multi-qubit gates, individual addressing is achieved by choosing the
frequency of the rf field. Corrections are small under the condition
$\Omega_g\ll \Delta E_{Z}/\hbar$ where $\Delta E_{Z}$ is the
difference in Zeeman shifts between two neighbouring ions. An
important advantage over the regular quantum computing scheme is
that the carrier and the sideband can be chosen by interference and
not by setting the frequency on resonance with one of these
transitions; i.e., the choice is set by the phase difference between the two rf drives.
This enables fast gates which are not limited to $\Omega_g \ll \nu$ but by $\eta \Omega_g \ll \nu$.

In conclusion, we have shown that by using a dressed four level
system, qubits can be realized that are resilient to magnetic field
noise which otherwise imposes a barrier for the coherence time in
various implementations of QIP. Still, the dressed states'
ability to support microwave-based conditional quantum dynamics is preserved.
Detailed experimental investigations
of the preparation and detection of microwave-dressed states have
been conducted, and measurements of the lifetime of dressed states
and of the coherence time of superpositions of dressed qubit states
are reported revealing an improvement compared to atomic states by
more than two orders of magnitude. It is shown that dressed states
allow for fast multi-qubit gates even in the
presence of a small effective Lamb-Dicke parameter, and that scaling
to arrays of trapped ions is possible. The insight gained in this
work removes a major obstacle for laser-free quantum information
processing (but is also applicable to laser-based schemes). 
Furthermore, this scheme is not restricted to trapped
ions and is in fact applicable to other physical systems where
dephasing due to external perturbations plays a role, for instance,
neutral atoms \cite{Specht2011} and solid state systems such as NV-centres in diamond
\cite{Simon2010}, ion-doped crystals \cite{Simon2010}, and
circuit-QED \cite{Clarke2008}.

\section{Methods}
\subsection{The experimental system}
Transitions between ground state hyperfine levels of a single $ ^{171}$Yb$^{+}$ ion
confined in a miniature
Paul trap (diameter of 2 mm) are driven with microwave radiation close to
12.64 GHz \cite{Hannemann2002} which is generated by mixing the signal from
a fixed frequency source at 6.3 GHz with an rf signal whose
frequency, amplitude, and phase are adjustable. For
STIRAP, two microwave fields are used with Gaussian amplitude
envelopes shifted in time relative to each other driving the
$\ket{-1} \leftrightarrow \ket{0}$ and the $\ket{+1} \leftrightarrow
\ket{0}$ transitions, respectively (see Fig. 1)

$^{171}$Yb$^+$ is produced from its neutral precursor by
photoionization using a diode laser operating near 399 nm.
Laser light near 369 nm driving resonantly the
S$_{1/2}$ F = 1 $\leftrightarrow$ P$_{1/2}$ F = 0 transition in
Yb$^+$ is supplied by a frequency doubled Ti:Sa laser, and serves
for cooling and state selectively detecting the ion. Initialization
in the state S$_{1/2}$ F = 0 ($| 0 \rangle$) is done using 369 nm
light tuned to the S$_{1/2}$ F = 1 $\rightarrow$ P$_{1/2}$ F = 1
transition.
A diode laser delivers light near 935 nm and
drives the D$_{3/2}$ $\leftrightarrow$ $[3/2]_{1/2}$ $~$
transition to avoid optical pumping into the metastable D$_{3/2}$
state during the cooling and detection periods.

The effectiveness of the STIRAP process for creating dressed states
and transferring them into the final state has been investigated as
a function of the parameters characterizing the pulse sequence (compare
\cite{Sorensen2006} for optical STIRAP). The sequence is divided into
discrete time increments $\Delta t = 1/(f_{\Omega}N_t)$ with a
positive integer $N_t$ and $f_{\Omega}= \Omega/(2\pi)$ where
$\Omega$ is the peak Rabi frequency. The width of the Gaussian
pulses is given by $N/f_{\Omega}$, with an integer number $N$. The
STIRAP sequence is robust against  variations of these parameters:
we investigated the range $10 \leq N_t\leq 40$ for fixed $N=10$ and
found no variation in the effectiveness of the pulse sequence, that is, the overall fidelity
of initial preparation of $\ket{0}$, preparation of  $\ket{D}$, and read-out stays constant 
(within statistical variations) at a value of about 93\% 

When
measuring sequences with pulse widths varying in the range $2 \leq
N\leq 20$ equally good results were obtained for $N\geq 4$. The
separation in time $s_t/f_{\Omega}$ of the two Gaussian pulses was
varied over the range $0\leq s_t/f_{\Omega} \leq 40/f_{\Omega}$, for
a pulse width where $N = 10$ to obtain a plateau of high
effectiveness ($\approx 93$\%) for the number $s_t$ belonging to the interval $10\leq
s_t \leq 20$. The best performance is obtained for equal detunings $\Delta_+ = \Delta_-$, ideally $\Delta_+ =0 = \Delta_-$, and for small
relative detunings. 
An experimental investigation yielded for  $\left|\delta\right| = \left| \Delta_+ - \Delta_- \right| <0.1 \Omega$ no statistically significant variation of the dressed state preparation fidelity.

In order to prepare, for example, the $\ket D$ state, the population
is first transferred from the initial state $\ket 0$ to the atomic
state $\ket{-1}$ by applying a microwave $\pi$-pulse. The first
half of the STIRAP sequence  then transfers the atomic population to
the dressed state $\ket D$.

A hold in the evolution of the two STIRAP microwave fields is
introduced at the crossing point of the amplitude envelopes of the
two pulses. At this point in time the occupation of the $\ket D$
state is at its maximum. Any dephasing of the $\ket D$ state or
transitions to other states during the holding time $T$ gives rise
to imperfect population transfer during the second half of the
STIRAP sequence that transfers the system to the atomic state
$\ket{+1}$.

Probing the state after the complete STIRAP sequence does not
distinguish between $\ket{-1}$ and $\ket{+1}$, as both yield
bright results upon final detection of resonance fluorescence.
Therefore, a $\pi$-pulse swaps the population of $\ket{+1}$ and
$\ket{0}$ before final detection. Thus, a dark result indicates a
successful STIRAP transfer between atomic states and dressed states
and a lifetime $T$ of the dressed state prepared during the the
holding period. The effectiveness of the STIRAP sequence as a
function of  the holding time $T$ is shown in Fig. 2.

In order to record Rabi oscillations between dressed states we first
prepare state $\ket{B}$, then an rf pulse in resonance with
$\ket{0'}\longleftrightarrow\ket{\pm 1}$ (Fig. 1a)  is applied during the
hold time $T$ in the evolution of STIRAP. This rf-pulse induces Rabi
oscillations between $\ket{B}$ and $\ket{0'}.$  After the Rabi
pulse, the STIRAP pulse sequence is completed and the population of
the atomic state $\ket{0}$ is probed as described above. If during
the rf Rabi pulse state $\ket{0'}$ is populated, then the second
part of the STIRAP sequence has no effect, that is, $\ket{0}$ is not
populated at the end of the experimental sequence.

\subsection{Theory: Single-qubit gate}
Starting with the Hamiltonian
\bea
H_{sqg}=&& \omega _{0}\ket 0  \bra 0
+\lambda_0(\ket {1} \bra {1} -
\ket {-1} \bra {-1}) + \nonumber \\
&&\Omega \left( \ket {-1} \bra {0}
e^{-i\omega _{-1}t}+\ket {1} \bra {0}
e^{-i\omega _{1}t}+h.c\right)+ \nonumber\\
&&\Omega_g \left( \ket {-1} \bra {0'}
e^{-i\lambda_0 t}-\ket {1} \bra {0'}
e^{i\lambda_0 t}+h.c\right)
\eea
if we set $\omega_{-1}=\omega_0 + \lambda_0,$ $\omega_{1}= \omega_0 - \lambda_0,$
moving to the interaction picture with respect to the time independent part  we get:
\bea
H^I_{sqg}=
&&\sqrt{2}\Omega \left( \ket {B} \bra {0}
+h.c\right)+\sqrt{2}\Omega_g \left( \ket D  \bra {0'}
+h.c\right)\nonumber \\
=&&\frac{\Omega}{\sqrt{2}}\ket{u}\bra{u}-\frac{\Omega}{\sqrt{2}}\ket{d}\bra{d}+\sqrt{2}\Omega_g
\left( \ket D  \bra {0'} +h.c\right) \eea The first two terms shift states $\ket u$ and $\ket d$ (that both contain $\ket B$) away from the $\ket D$, and the second part creates the
gate. The interactions created by the two microwave fields with Rabi frequency $\Omega$ and the rf field $\Omega_g$
shown in Fig. 1a permit the implementation of this Single-qubit
gate. The two rf fields have a $\pi$ phase difference at the initial time of the pulse which realizes the coupling to the $\ket D$ state and not the $\ket B$ state.

\subsection{Theory: Multi-qubit gate}
The levels and fields of Fig. 1b are described by the
Hamiltonian:

\bea H^I_{mqg}=&& \omega _{0}\ket 0 \bra 0
+\lambda_0(\ket {1} \bra {1} -
\ket {-1} \bra {-1}) + \nonumber \\
&&\Omega \left( \ket {-1} \bra {0}
e^{-i\omega _{-1}t}+\ket {1} \bra {0}
e^{-i\omega _{1}t}+h.c\right)+ \nonumber\\
&&\Omega_g \left( \ket {-1} \bra {0'}
e^{-i(\lambda_0 -\delta)t}+\ket {1} \bra {0'}
e^{i(\lambda_0 - \delta)t}+h.c\right)\nonumber\\
&&+ \nu b^{+}b
_{1}\ket {1} \bra {1}  + \lambda \left(
\ket {-1} \bra {-1} -\ket 1 \bra {1} \right) \left( b+b^{+}\right) , \eea
where $b$ and $b^+$ are the annihilation and creation operators of
the trap and $\lambda$ is proportional to the magnetic gradient as
in \cite{Mintert2001}. In the interaction picture with respect to
the microwaves and after the Polaron transformation: \\$U=e^{\eta
\ket {+1}  \bra {+1} (b^{+}-b)}e^{-\eta \ket {-1} \bra {-1}
(b^{+}-b)}$ we get:\\ $UHU^{+}=\nu b^{+}b +\Omega _g\left( \left\vert
-1\right\rangle \left\langle 0'\right\vert e^{-\eta
(b^{+}-b)}e^{i\delta t}+\left\vert 1\right\rangle \left\langle
0'\right\vert e^{\eta (b^{+}-b)}e^{i\delta t}+h.c\right).$
In first order in the Lamb
Dicke parameter we get: \be \sqrt{2}\eta \Omega _g\left( \left\vert
D\right\rangle \left\langle 0'\right\vert e^{i\delta t}-h.c\right)
(b^{+}-b) \ee   
In zeroth order we get a coupling between the $\ket B$ and $\ket {0'},$ this term is ignored since the coupling $\Omega_g$ is 
much smaller then the energy gap which is of the order $\Omega$. 
Note however, that by changing the phase between the rf fields a gate with
respect to the $\ket B$ state can be realized.

\bigskip

\section{Acknowledgments}

We acknowledge support by the Bundesministerium f\"{u}r Bildung und
Forschung (FK 01BQ1012), Deutsche Forschungsgemeinschaft, European
Commission under the STREPs PICC and GIF, secunet AG, and the
Alexander von Humboldt Foundation. Technical help with the microwave
set-up by T. F. Gloger is acknowledged.


\end{document}